\documentclass[12pt,aps,prb,preprint]{revtex4}   % style for Physical Review B and AJP are similar

\usepackage{amsmath}    % need for subequations
\usepackage{graphicx}   % for figures

\begin{document}

\title{A better presentation of Planck's radiation law}	% shortened by dvs
\author{Jonathan M.\ Marr}
\email{marrj@union.edu}
\author{Francis P.\ Wilkin}
\email{wilkinf@union.edu}
\affiliation{Union College, Schenectady, NY}

\date{\today}

\begin{abstract}
%A common discussion in introductory physics and astronomy courses uses Wien's displacement law to explain the colors of blackbodies in terms of their temperatures.  Sometimes this is used to address the color of the Sun and stars.  
Introductory physics and astronomy courses commonly use Wien's displacement law to explain the colors of blackbodies, including the Sun and other stars, in terms of their temperatures.
We argue here that focusing on the peak of the blackbody spectrum is misleading for three reasons.  First, the Planck curve is too broad for an individual spectral color to stand out. Second, the location of the peak of the Planck curve depends on the choice of the independent variable in the plot.  And third, Wien's displacement law is seldom used in actual practice to find a temperature and direct fitting to the Planck function is preferable.  We discuss these flaws and argue that, at the introductory level, presentation of blackbody radiation in terms of photon statistics would be more effective pedagogically.  The average energy of the emitted photons would then be presented in place of Wien's displacement law, and discussion of the Stefan-Boltzmann law would include the total number of photons emitted per second.  Finally, we suggest that the Planck spectrum is most appropriately plotted as a ``spectral energy density per fractional bandwidth distribution,'' using a logarithmic scale for the wavelength or frequency.
%that is $\lambda B_\lambda$ vs.\ ln($\lambda$) or $\nu B_\nu$ vs.\ ln($\nu$).
\end{abstract}

\maketitle

\section{Introduction}

The concept of blackbody radiation, along with the associated Stefan-Boltzmann law and Wien's displacement law, is a crucial pillar of physics and astronomy.
%so fundamentally important to physics and astronomy that this complex topic is viewed as a crucial pillar to be covered in all undergraduate curricula.  
In introductory courses these laws are applied to the cosmic background radiation and to stars.  Unfortunately, however, there are three significant flaws in conventional presentations which lead to misconceptions.  

First, Wien's displacement law is overinterpreted.  
The idea that a star's color is given by the location of the peak of its Planck curve requires that the amount of energy emitted in that spectral color be significantly greater than in the other colors.  In fact, though, the Planck curve is so broad and the peak so gradual that across the small width of the visible band the spectrum near the peak is effectively flat.

Second, the plotting of a Planck curve and an expression for Wien's displacement law involve a necessary choice of independent variable.  In most introductory physics and astronomy classes, the Planck curve is plotted as a function of wavelength.
More precisely, the formula plotted is
$B_\lambda$ vs.\ $\lambda$, where $B_\lambda$ is the emitted power per unit area per steradian per wavelength interval and is given by
\begin{equation}
\label{Blam}
B_\lambda =\frac{2hc^2}{\lambda^5}\frac{1}{\exp[hc/(\lambda kT)]-1},
\end{equation}
where $h$ and $k$ are the Planck and Boltzmann constants and $c$ is the speed of light.  (The equation itself is often omitted in introductory courses, but the curve is shown in a figure.)
However, an equally correct Planck curve, often used in more advanced courses, is a plot of emitted power per unit area per steradian per 
{\it frequency} interval.  This function is denoted $B_\nu$, and is given as a function of frequency $\nu$ by
\begin{equation}
\label{Bnu}
B_\nu =\frac{2h\nu^3}{c^2}\frac{1}{\exp[h\nu/(kT)]-1}.
\end{equation}
%In some physics texts, the Planck law is defined as a spectral energy density, which is identical to $B_\lambda$ or $B_\nu$ up to a constant factor.  [XX I don't think this sentence is necessary --dvs. XX]
The functions $B_\nu$ and $B_\lambda$ describe the same physics, but they have different shapes, due to the nonlinear change of variable from wavelength to frequency.  As a consequence, the two curves peak at different locations in the spectrum.  Unfortunately, though, most presentations of Planck's law fail to acknowledge the subjective choice of how to plot the spectrum, and this affects students' ability to interpret the curve, and the location of its peak, correctly.
%understanding of the true nature of Planck's curve, including Wien's displacement law. 

Third, despite its usual application in textbooks,  
Wien's displacement law is not generally used for determining the temperature of a thermal source in real scientific research.  
Wien's displacement law is most useful for roughly predicting the spectral region in which a thermal source will radiate most intensely.  But to obtain a numerical value of the temperature, scientists generally 
adjust the temperature in the equation for the Planck function to fit measurements of intensity.
%the entire Planck function, fitting this function to two or more measurements of the spectrum. 
 
Attention to the misunderstanding of Wien's displacement law has been raised numerous times, as early as 1954, in the journals of many different disciplines, from general physics to optics, thermal physics, astronomy, and 
engineering.\cite{Bracewell,Ross,Chiu,Brecher,LS,SL,Over,Heald,ZW,Stewart}  Recognition of this issue, however, has not taken hold in introductory textbooks.  We suspect that these previous discussions, although enlightening, have left readers unsure of how to improve upon the standard presentation.  In this paper, we propose a method for correcting the presentation which, we hope, will facilitate the needed change in the pedagogy. 

%In our experience in teaching undergraduates, we have also found that Planck's law is one topic which students struggle to comprehend.  The conventional approach in teaching this material, therefore, does not succeed very well in conveying the important principles involved.  [XX I'm not sure this vague, anecdotal remark adds much to the introduction. XX]

In the next section we discuss these flaws in greater detail and then in Section III we propose an alternative approach for presenting Planck's law which avoids these conceptual difficulties.  

\section{Three strikes against teaching Wien's displacement law}

\subsection{The Sun is white.}

To most people it is a ``fact'' that the Sun is yellow.  The explanation of this ``fact'' by way of Wien's displacement law is an all too-tempting exercise for students and young instructors.  Unfortunately, the conventional approach for discussing Planck's law in introductory physics and astronomy texts mostly reinforces the idea that the color of the Sun can be explained by Wien's displacement law.  A typical discussion starts with the statement that Wien's displacement law shows that as the temperature of a blackbody increases its color shifts blueward in the spectrum, and then the surface temperature of the Sun is used for inserting some real numbers.  At the Sun's surface temperature of 5800 K, the wavelength of the peak of $B_\lambda$ occurs at approximately 500 nm, which, as most texts then state, is in the middle of the visible band.  (Some texts, mistakenly, then conclude that the Sun is actually green but appears yellow because of the atmospheric scattering and/or the complex issues of color vision.\cite{CK,IH})  Although many texts correctly explain that the Sun is white but appears yellow when at lower elevations because of the greater scattering of shorter wavelengths by the atmosphere,\cite{PF} this is still, unfortunately, an over-simplified presentation which leaves the students with a false sense of the usefulness of Wien's displacement law. 

In reality, the apparent color of a star cannot simply be determined by the spectral color corresponding to the peak of that star's blackbody spectrum.  
%The sensitivity of the human eye to light is a logarithmic dependence and so [XX I think it's a non-sequitur to bring up "logarithmic" here, since the dependence could be logarithmic but still very sensitive to small ratios. XX]
The response of the human eye to light has a logarithmic dependence and so the {\it apparent} relative brightnesses in the amounts of light in the different spectral colors are given by the {\it ratios} of the intensities, not the absolute differences.  Near the peak of the Planck curve, these ratios will all be close to one.  To the human eye, then, the variation in the amounts of the different spectral colors in the Sun's radiation is, actually, barely noticeable.  Blackbodies of slightly different temperatures do appear to have slightly different hues, but assigning an individual spectral color to a star by calculating the peak of its Planck function is inappropriate.

To make this point quantitatively, we have converted $B_\lambda$ into a function similar to the stellar magnitude scale, a familiar scale to many naked-eye visual observers.  Stellar magnitudes are logarithmically related to flux and are defined by 
\begin{equation}
m_1 - m_2 = -2.5 \log({F_1 \over F_2}),
\end{equation}
where $F_1$ and $F_2$ are the fluxes from stars 1 and 2 and $m_1$ and $m_2$ are the {\it magnitudes} of stars 1 and 2.  (Note that a brighter star has a smaller magnitude and stars brighter than Vega, the calibration standard, will have negative magnitudes.)  The brightest star in the night sky, Sirius, has a magnitude of $-1.5$ while the faintest stars visible with the naked eye on a moonless night at a location with no artificial lights are about magnitude~6.
We have similarly devised a ``relative-magnitude'' spectrum by calculating a wavelength-dependent logarithmic brightness given by
\begin{equation}
m(\lambda)- m(\lambda_\textrm{ref}) = -2.5  \log({B_\lambda(\lambda) \over B_\lambda(\lambda_\textrm{ref})}),
\end{equation}
where $B_\lambda$ is given by Eq.~(\ref{Blam}) and the zero-point is set at some $\lambda_\textrm{ref}$.\cite{Vega}
Figure~\ref{ThreeStarsB} shows the relative-magnitude spectrum for the Planck curve at three different temperatures.  The solid, black curve corresponds to 
$B_\lambda$ vs.\ $\lambda$ at the Sun's surface temperature.  Near the peak, the variation of the relative magnitude over 50 nm in wavelength, the approximate width of a color band in the ROYGBIV rainbow, 
%[XX WHAT DO YOU MEAN BY THE WIDTH OF A SPECTRAL COLOR? DO YOU MEAN THAT THE EYE CAN'T DISTINGUISH SPECTRAL COLORS SEPARATED BY LESS THAN THIS, OR MERELY THAT OUR MOST COMMON COLOR NAMES ARE USED TO COVER SPECTRAL BANDS THAT ARE ABOUT THIS WIDE? XX]
is less than 0.07 magnitudes. 
The greatest difference across the entire visible band, between the peak and the red edge (at 700 nm), where the eye's sensitivity is greatly diminished, is still only about 0.6 magnitudes.  Considering that skilled amateur astronomers can judge relative brightnesses, without regard to differentiating color, as small as 0.1 magnitudes, these differences are too small to cause the Sun to appear to be of a single spectral color, be it green or yellow.  Therefore, using Wien's displacement law to address the color of the Sun is, actually, an overinterpretation of its power.  In reality, the peak is very broad compared to the wavelength range of a spectral color, and with the human eye's logarithmic response all the colors in the visible window appear of comparable brightness.

Of course, stars of more extreme temperatures do have more readily apparent colors; the coolest stars appear reddish-orange and the hottest stars are blue-white.  The perception of these colors, though, does not conflict with our discussion here that the Sun is white.  Also shown in Fig.\ \ref{ThreeStarsB} are curves corresponding to $B_\lambda$ vs.\ $\lambda$ for blackbodies at the temperatures of 30,000~K and 4000~K, temperatures characteristic of hot and cool stars, respectively.  These curves demonstrate that the colors of these stars result not because the peaks of their Planck functions occur at the wavelengths corresponding to red and blue, but because their spectra are significantly sloped across the visible band.  The relative magnitude spectra of these stars are seen to differ by 2.6 and 4.8 magnitudes, respectively, from one end of the visible spectrum to the other.  We see, therefore, that the coolest and hottest stars do have significant differences in the power radiated at different colors while stars with moderate temperatures (such as the Sun) are, effectively, white, with only slight differences in hue.  

\subsection{$B_\lambda$ vs.\ $B_\nu$ and the choice of independent variable}

A more significant problem with using the peak of $B_\lambda$ to define a ``peak color'' involves the subjective choice of the independent variable in the plot.  A common point of confusion, even among Ph.D.\ physicists, arises from the fact that the two standard forms of the Planck function, 
$B_\lambda$ vs.\ $\lambda$ [see Eq.~(\ref{Blam})] and 
$B_\nu$ vs.\ $\nu$ [see Eq.~(\ref{Bnu})], peak at different wavelengths.  At the Sun's surface temperature, for example, $B_\nu$ peaks at a wavelength of 880 nm, which is in the infrared.  How could this be?  Isn't the Planck function defined well enough that regardless of how we plot it we should come to the same qualitative conclusions about the source?
The answer lies in the fact that $B_\lambda$ and $B_\nu$ are not, actually, the same function.  If one substitutes $\lambda = c / \nu$ into Eq.~(\ref{Blam}) one does not obtain the expression for $B_\nu$ in Eq.~(\ref{Bnu}).
The difference between these functions revolves around the method by which the spectrum is determined.  The former results from distributing the radiation into equal bins of wavelength and the latter into equal bins of frequency.  There is nothing more fundamental about analyzing the spectrum in the wavelength domain than in the frequency domain.  Furthermore, as commented by other authors, the Planck function can be plotted with an assortment of choices of independent variables:  The independent variable can also be chosen to be $\nu^2$ (which approximately mimics dispersion by a prism) or $\ln\nu$, for example.\cite{Heald,Stewart}  

However, any change in the independent variable requires a corresponding change in the functional form of the spectrum such that the integrated power is preserved. That is,
the integrals of each function over any defined range of the spectrum must agree, for example,
\begin{equation}
\label{integB}
\int_{\nu_1}^{\nu_2}{B_\nu \,d\nu}=\int_{\lambda_2}^{\lambda_1}{B_\lambda \,d\lambda}, 
\end{equation}
where $\nu_1 = c / \lambda_1$ and $\nu_2 = c / \lambda_2$, so that they describe the same distribution of emitted power throughout the spectrum.  For each choice of the independent variable, there is a corresponding spectral peak location.  

Any of these spectral functions is a correct physical description of blackbody radiation, but their shapes differ because of 
the nonlinear relations between the different independent variables.  Considering the traditional independent variables $\lambda$ and $\nu$, we have
\begin{equation}
\label{dnuvdlambda}
|d\nu| = {\nu \over \lambda} |d\lambda|.
\end{equation}
The effect of this nonlinear relation on the shape of the curve is two-fold.  First, the horizontal-axis steps, when comparing the two plots, are skewed.  As demonstrated visually in the paper by Soffer and Lynch,\cite{SL} equal steps of $\Delta\lambda$ in the 
$B_\lambda$ vs.\ $\lambda$ plot correspond to steps of $\Delta\nu$ in the $B_\nu$ vs.\ $\nu$ plot that are stretched at the higher-frequency end and compressed at the lower frequencies.  Second, as required by Eq.~(\ref{integB}), the functions on the vertical axes must differ to compensate for the unequal steps along the horizontal axes.  The vertical-axis values of one plot are increased relative to the other plot at one end and decreased at the other end.  

This apparent disagreement between different Planck curves, actually, has nothing to do with the physics behind the Planck function itself.  Consider, for example, a source whose spectrum in the visible window
when plotted as power per unit area per steradian per frequency interval is inversely proportional to frequency, so that it decreases toward the blue end, with the functional form
\begin{equation}
I_\nu = K \nu^{-1},
\label{I0}
\end{equation}
where $K$ is a constant.  But, when calculated per {\it wavelength} interval, the same spectrum is
\begin{equation}
I_\lambda = K \nu^{-1} \left|{d\nu \over d\lambda}\right| = K \lambda^{-1},
\label{Ilambda}
\end{equation}
which decreases toward the red end.  Expressions (\ref{I0}) and (\ref{Ilambda}) do not disagree, because when integrated they give the same amount of energy radiated over a given region of the spectrum.  However, a visual representation of either function can be misleading.  Many young scientists could easily be deceived about a basic aspect of the spectrum such as whether the source
is brighter in the blue or in the red.  We see, therefore, that the fundamental issue here is really the choice of how to plot spectra.  Soffer and Lynch\cite{SL} and Stewart\cite{Stewart} point out that the intensity of the emitted radiation is, in fact, a distribution function, which by definition, has a shape that depends on the choice of independent variable.

A blackbody source of given temperature, $T$, radiates a total power (per area per steradian) across the entire electromagnetic spectrum of $\sigma T^4 / \pi$, where $\sigma$ is the Stefan-Boltzmann constant. (The factor of $1 / \pi$ does not appear in the Stefan-Boltzmann law due to the integration of intensity over solid angle.\cite{Gray})  This formula is obtained by integrating $B_\nu$ over all frequencies, from zero to infinity, that is,
\begin{equation}
\label{intBnu}
\int_{0}^{\infty}{B_\nu \,d\nu} = {\sigma T^4 \over \pi}.
\end{equation}
Similarly, one can integrate from $\nu$ = 0 to $\nu_0$ to
determine the total power (per area per steradian) emitted at all frequencies below $\nu_0$.
One could also calculate the total power (per area per steradian) over this spectral range by integrating $B_\lambda$ over wavelength.  Provided that the integrals start from the same end of the spectrum, for a given temperature, T, these integrations will yield the same value, and so
\begin{equation}
\label{PlambdaEqPnu}
\int_{0}^{\nu_0}{B_\nu \,d\nu} = \int_{\lambda_0}^{\infty}{B_\lambda \,d\lambda},
\end{equation}
where $\lambda_0 = c / \nu_0$.  The function represented by this integration is the
{\it cumulative distribution function} of the radiated power of a blackbody source of given temperature.  It describes the total power emitted from one end of the spectrum up to any particular point and its value at any given
point is the same regardless of the independent variable used in the spectrum.  
We have, therefore, a function that is independent of the choice of independent variable and so we can also denote it as $P(X)$, where $X$ is any variable that can be used to indicate location in the spectrum.  
For convenience we will use the often-used unitless variable $x$ defined by
\begin{equation}
\label{xeq}
x = \frac{h\nu}{kT} = \frac{hc}{\lambda kT}.
\end{equation}
Then, for any given temperature, T, $P(x)$ is given by
\begin{equation}
\label{Pofx}
P(x) = {2k^4T^4 \over h^3c^2}\int_{0}^x{{t^3\,dt \over e^{t}-1}}.
\end{equation}
Unfortunately, we cannot write a straightforward expression for $P(x)$ since
the integral in Eq.~(\ref{Pofx}) is not expressible in closed form in terms of elementary functions, although it can easily be evaluated numerically.\cite{Pavelle,Stewart}
Of course, a cumulative distribution function depends on the choice of the end of the spectrum at which the integration starts; here, we have started from the low frequency end, but one could just as easily start from the low wavelength end.  

Given this cumulative distribution function, the Planck function $B_\nu$ is obtained by 
\begin{equation}
\label{dPdnu}
B_\nu = {dP(x) \over dx} {dx \over d\nu}.
\end{equation}
Similarly, for any choice of independent variable, the relation between the corresponding Planck function and $P(x)$ is obtained by replacing $\nu$ in Eq.~(\ref{dPdnu}) by the new independent variable.  For example, $B_\lambda$ is obtained by taking the derivative with respect to wavelength,
\begin{equation}
\label{dPxdlam}
B_\lambda = {dP(x) \over dx} {dx \over d\lambda}.
\end{equation}

We now relate these considerations to Wien's displacement law.  The peak of the Planck function in any form is determined by finding where its derivative equals zero, and we now see that the choice of Planck function depends on the choice of variable of differentiation of $P(x)$.  The location of the peak, therefore, is merely the zero point of the {\it second derivative} of the cumulative distribution function.  It should not, therefore, be a surprise that different choices of the differential variable lead to different locations of the zero in the second derivative.  Ross\cite{Ross} and Stewart\cite{Stewart} conclude that the designation of any peak of the function is not meaningful and should, therefore, be de-emphasized.  

On the other hand, despite the ambiguity posed by Wien's displacement law, the general concept that hotter blackbodies emit photons primarily of higher energies is certainly correct and important to impress upon physics and astronomy students.  Heald\cite{Heald} and Stewart\cite{Stewart} suggest as an alternative ``Wien peak'' that we use the frequency below which half the emitted power is contained.  In this paper, we propose that the average energy of the emitted photons be used.  The average photon energy both serves the pedagogical goals better (as we explain below) and has a clear physical meaning.

With regards to the choice of how to present the Planck curve, we also argue below that the logical approach is to use $\ln\lambda$ or $\ln\nu$ as the independent variable, as proposed by Bracewell\cite{Bracewell} and by Zhang and Wang.\cite{ZW} 

\subsection{Wien's displacement law is not used for determining temperatures.}

The students in an introductory course naturally assume that the instructor presents material that teaches them how physics is done, i.e., that they are taught methods used in physics research.   When they are introduced to Wien's displacement law, with examples that relate the temperature of the radiating body to the peak of its Planck curve, they are given the impression that this is how the temperature of a star, for example, is obtained. In some texts, the use of Wien's displacement law as a means of determining a star's temperature is made quite explicitly.\cite{AS,CM,Hester}  The reality, though, is that Wien's displacement law is {\it not} used to determine a star's surface temperature.  The determination of stellar photospheric temperatures is accomplished through the ratio of spectral line intensities, when available, or the ratio of fluxes through different filters for rougher estimates, but never by fitting the peak of the spectrum.  Or, consider the determination of the temperature of the cosmic background radiation (and fluctuations from the average), which is given simply by fitting the observed intensity at any given frequency to the Planck function.  

In principle, the intensity of blackbody radiation at any given wavelength depends only on the body's temperature, and so a single intensity measurement is easily translated directly into a temperature.  And, if the source is unresolved, so that the intensity is not known, the temperature can be inferred by fitting the measurements of the flux in any two small bandpasses.  The ratio of the two fluxes removes the source angular size dependence, leaving the temperature as the only variable.  On the other hand, to use Wien's displacement law one needs to determine the peak location, and that requires fitting a curve to a number of flux measurements; even a rough fit with a peak requires at least three data points.  If one uses the Planck function to fit the data points to find the peak, then one is already obtaining a measure of the temperature before learning of the peak location.  And, if the peak is obtained without using the Planck function, then the inferred peak location is dependent on the assumed curve shape and on the wavelengths at which the measurements were made, unless many measurements are made at closely spaced wavelengths and with very small flux uncertainties.  With three flux measurements, one can more easily just use two different flux ratios, to yield two independent estimates of the temperature.  In short, there is no situation where a measure of the location of the peak of the Planck function yields a better estimate than either fitting an intensity directly to the Planck function or fitting the ratio of fluxes.  

Wien's displacement law {\it is} useful, however, for identifying the general region of the entire electromagnetic spectrum in which a thermal source of given temperature will be brightest.  The wavelengths of the peak brightness for the relevant dispersion rules, from a $\nu^2$-dispersion rule to a $\lambda$- dispersion, all occur within half an order of magnitude of each other.\cite{Heald,Stewart}  In the search for detection of the cosmic background radiation, for example, the researchers needed to know that detectors in the microwave region were needed (and hence the alternative name ``cosmic microwave background'').  In this paper, though, we emphasize that at the introductory level the discussion of Wien's displacement law with examples relating the colors of stars to their surface temperatures leads to misunderstanding by students.

\section{Suggestions for how to present Planck's law}

\subsection{Discuss the average photon energy in place of Wien's displacement law.}

We have argued, following the discussions of other authors, that Wien's displacement law can be misleading---especially to beginning students.  Furthermore, in our experience, students with limited mathematical background are not prepared to fully understand such subtle concepts as the {\it peak of the curve} and {\it area under the curve}.  The discussion would be more concrete, and hence easier to grasp, if it were based on the more easily visualized emission of photons.  And, since the Planck curve is a statistical distribution, students would understand it better if it were discussed in more typical statistical terms.  The most easily understood statistical measure of a distribution is the mean.  We therefore propose that the concept that hotter blackbodies radiate higher energy photons, on average, be conveyed by presenting the \emph{average energy of the emitted photons} instead of Wien's displacement law.

Instructors may feel the desire to teach about blackbody radiation in a way that adheres to the historical development.  This is an admirable goal, in general, but in this particular instance it may not be pedagogically preferable.  Although the photon was discovered slightly after the laws of blackbody radiation, we argue that the radiation laws would be conveyed more effectively if introduced in terms of photons.  It is significantly easier for undergraduates to comprehend the meaning of the average photon energy than the meaning of the peak of the Planck function.  Perhaps if Wien had known of photons at the time, he would have expressed his law in these terms.  

The average energy of the photons emitted by a blackbody of temperature $T$ is given by
\begin{equation}
\label{enavg}
\langle E_{\textrm{phot}}\rangle = {\textrm{Total~energy~emitted}\over \textrm{Number~of~photons}} = \left( \int_0^\infty{B_\nu(T)\, d\nu} \over \int_0^\infty{[B_\nu(T)/{h\nu}]\, d\nu}\right).
\end{equation}
The integral in the denominator yields
($30 \zeta (3) /\pi^5 k)\, \sigma T^3$, where 
$\sigma = {2 \pi^5 k^4 /(15 c^2 h^3)}$ and $\zeta(3)$, which equals 1.2021, is the Riemann Zeta function with argument~3. The numerator in Eq.~(\ref{enavg}) is given by Eq.~(\ref{intBnu}), so 
the average photon energy reduces to
\begin{equation}
\label{Eavg}
\langle E_{\textrm{phot}}\rangle = {\pi^4 \over {30 \zeta (3)}} kT = (3.7294\times 10^{-23}~\textrm{J/K})\cdot T.
\end{equation}
For comparison with Wien's displacement law, the wavelength corresponding to the average photon energy is given by
\begin{equation}
\label{lambdaavg}
\lambda_{\langle E\rangle} T = 0.53265~{\textrm{cm}\cdot\textrm{K}}.
\end{equation}
The average-energy photons emitted from the Sun's surface, for example, have wavelength 920 nm.

\subsection{Discuss the number of photons as part of the Stefan-Boltzmann law.}

Misconceptions about the Stefan-Boltzmann law are also common.  Even students who understand that a hotter body radiates photons of higher energy, on average, sometimes fail to recognize that the increase in power is mostly due to an increase in the number of photons emitted per second. 

Discussion of blackbody radiation in terms of photons, rather than just in terms of energy, can also be extended to the Stefan-Boltzmann law, which can be presented in two steps.  First, instructors can give the total number of photons emitted per second per unit area, as a function of temperature.  
We first remind the reader that the energy flux (the Stefan-Boltzmann law) is given by 
$F = \sigma T^4 = \pi \int_0^\infty{B_\nu\, d\nu}$.  The photon flux, $N_{\textrm{phot}}$, similarly, is given by 
\begin{equation}
\label{Nphotons}
N_{{\textrm{phot}}} = \pi \int_0^\infty{{B_\nu \over h\nu}\, d\nu}=
(1.5205 \times 10^{15}~{\rm photons~m^{-2}~s^{-1}~K^{-3}})\cdot T^3 .
\end{equation}
And then, instructors can present the total power emitted per unit area, by simply combining Eqs.\ (\ref{Eavg}) and~(\ref{Nphotons}) to get
\begin{equation}
\label{SBphot}
F =  N_{\textrm{phot}} \langle E_{\textrm{phot}}\rangle = (5.6704 \times 10^{-8}~ {\rm J~m^{-2}~s^{-1}~K^{-4}})\cdot T^4,
\end{equation}
which is the Stefan-Boltzmann law.  

Note that this approach also spells out the two main concepts behind the Stefan-Boltzmann law more explicitly.  The students can see quite clearly in 
Eq.~(\ref{SBphot}) the two factors in the total radiated power---the average photon energy and the total number of photons emitted per second.

\subsection{Present Planck's law as a spectral energy per fractional bandwidth distribution.}

As discussed above, there are numerous choices for the independent variable in presenting a plot of Planck's curve, and the different choices yield differently shaped curves.\cite{Chiu,SL,Heald,Stewart}  The ambiguity is not in the thermal radiation physics but in the choice of how to plot spectra.
Although we propose in the previous sections that blackbody radiation be introduced in terms of photons, it is still desirable to describe the Planck curve in terms of the distribution of emitted energy, rather than the number of photons, since studies of radiation across the entire electromagnetic spectrum generally involve energy plots.

We present here the case for using $\ln\lambda$ or $\ln\nu$ as the independent variable.  In some astrophysical fields, presenting spectra in which the independent variable is logarithmic is already a common practice.\cite{ALS,Elv} This type of plot is often called the ``spectral energy distribution'' (or SED).\cite{CO}  Unfortunately, though, the definition of the term ``spectral energy distribution'' is not consistent in all fields.  Sometimes this term is used in the literature synonymously with ``spectrum'' (i.e.\ intensity vs.\ wavelength or vs.\ frequency).\cite{SED}  To circumscribe this ambiguity, we will refer to a spectral plot in which the independent variable is logarithmic as a ``spectral energy per fractional bandwidth distribution,'' and to indicate that this is a particular type of SED, we will abbreviate it as ``FBSED.''

For analysis of astronomical observations, in which one measures the flux density of the radiation, expressed as $F_\lambda$ (power per unit area per wavelength interval) or $F_\nu$  (power per unit area per frequency interval), one plots the FBSED as the quantity 
$\lambda F_\lambda$ vs.\ $\ln\lambda$, or $\nu F_\nu$ vs.\ $\ln\nu$.  (To avoid the confusing issue of the logarithms of quantities with units, the parameters $\nu$ and $\lambda$ can be divided by 1 Hz and 1 nm, or whatever units are most convenient.)  In theoretical studies, one is more interested in the total radiated power (or luminosity) and so  
one might plot 
$\lambda L_\lambda$ vs.\ $\ln\lambda$, or $\nu L_\nu$ vs.\ $\ln\nu$, where $L_\lambda$ and $L_\nu$ are the luminosity per wavelength interval and per frequency interval.  This function can be used to describe the radiation emitted by any source, but here, we apply this function specifically to blackbody radiation sources. 

First, we show that $\lambda B_\lambda$ or $\nu B_\nu$ is the correct function on the vertical axis when the horizontal-axis parameter is logarithmic.  For any given choice of independent variable, the proper vertical-axis function is one that yields the same energy output when integrated.  Starting from Eq.~(\ref{integB}), then, we multiply the left hand side by $\nu / \nu$ and the right hand side by 
$\lambda / \lambda$.  Equation~(\ref{integB}) then becomes
\begin{equation}
\int_{\nu_1}^{\nu_2}{\nu B_\nu \frac{d\nu}{\nu}}=\int_{\lambda_2}^{\lambda_1}{\lambda B_\lambda \frac{d\lambda}{\lambda}},
\end{equation}
which can be rewritten as
\begin{equation}
\int_{\nu_1}^{\nu_2}{\nu B_\nu\, d(\ln\nu)}=\int_{\lambda_2}^{\lambda_1}{\lambda B_\lambda\, d(\ln\lambda)}.
\label{nuBnu}
\end{equation}
The integrands in Eq.~(\ref{nuBnu}) are indeed those used in FBSED plots.  The same result can be obtained formally by substituting $\ln\nu$ for $\lambda$ in Eq.~(\ref{dPxdlam}).

Bracewell,\cite{Bracewell} and Zhang and Wang\cite{ZW} noted that if the independent variable is chosen to be logarithmic ($\ln\nu$ and $\ln\lambda$), the associated Planck functions 
($B_{\ln\nu}$ vs.\ $\ln\nu$ and $B_{\ln\lambda}$ vs.\ $\ln\lambda$) peak at exactly the same place.  As commented by Stewart, though, this is merely a mathematical convenience and not representative of any physical significance.\cite{Stewart}  In fact, these plots have the same peaks because they are identical plots.  Recall the reasons for the disagreement between the 
$B_\nu$ vs.\ $\nu$ and $B_\lambda$ vs.\ $\lambda$ plots discussed in the Introduction.  With an FBSED plot we now have horizontal-axis steps of d($\ln\nu$) and d($\ln\lambda$) which, according to Eq.~(\ref{dnuvdlambda}), are related to each other by
\begin{equation}
d(\ln\nu) = {d\nu \over \nu} = -{d\lambda \over \lambda} = -d(\ln\lambda).
\end{equation}
The negative sign occurs because an increase in frequency corresponds to a decrease in wavelength.  Therefore, equal horizontal-axis steps in one plot correspond to equal horizontal-axis steps in the other.  And one can easily verify that, on the vertical axes, $\nu B_\nu$ and $\lambda B_\lambda$ are the same function by using Eqs.\~(\ref{Blam}) and (\ref{Bnu}) and substituting $\nu = c / \lambda$.  When showing the Planck function in this form, therefore, there is no ambiguity about which one to consider.

There is a stronger justification for using FBSED plots as the standard approach.
The crux of the FBSED plot is the $\ln\lambda$ or $\ln\nu$ on the horizontal axis.  The choice of how to assign the steps along the horizontal axis is, essentially, the same concept as picking the spectral resolution element.  Now, in general, the wavelength resolution of a spectrum is proportional to the wavelength (and similarly with frequency).  In common practice, one sees spectral displays that cover only a tiny fraction of the full electromagnetic spectrum, but one has an intuitive sense of what a reasonable resolution should be, and that sense involves a percentage.  When viewing a visible-wavelength spectrum, for example, a resolution of tens of nanometers would be considered to be quite poor.  But, for spectra at radio-wavelengths, of order one cm, a resolution as tiny as tens of nanometers would be remarkable.  Consider, for example, that a spectrograph's resolving power, R, is defined as\cite{BGO} 
\begin{equation}
R = {\lambda \over \Delta\lambda}.
\end{equation}
A reasonable choice of how finely to divide the spectrum would use wavelength steps that were proportional to $\lambda$, and likewise with frequencies.  In fact, with this concept in mind, both $B_\lambda$ vs.\ $\lambda$ and $B_\nu$ vs.\ $\nu$ displays over a wide spectral range are non-optimal. They both involve horizontal-axis steps that are much smaller than the resolution 
at one end and much larger at the other.  These plots differ from standard spectral plots because they cover such a large range of wavelength.  For spectra covering many orders of magnitude in either wavelength or frequency, a logical choice for the horizontal-axis resolution elements is one that involves equal {\it fractional} intervals.  An FBSED plot, therefore, displays spectra across the entire electromagnetic spectrum in a form in which all parts of the spectrum have comparably reasonable resolution. This is, essentially, the display that shows all parts of the entire spectrum with equal clarity.  We propose, here, that this become the common practice.  

In terms of the unitless variable $x$, as defined by Eq.~(\ref{xeq}), the Planckian FBSED is given by
\begin{equation}
\label{SEDva2}
\nu B_\nu = \lambda B_\lambda = \frac{2k^4T^4}{h^3c^2}\frac{x^4}{\exp(x)-1}.
\end{equation}
In Fig.~\ref{SEDplot} we display the Planckian FBSED curve with $\ln x$ as the independent variable.  Note that by using $x$, the temperature dependence appears only in the vertical axis which can then be scaled by the temperature.  We get, then, a single curve to describe blackbody radiation for all temperatures.  The value of $x$ corresponding to the average photon energy is given by
\begin{equation}
\label{avgx}
x_{\langle E\rangle}= 2.7012,
\end{equation}
and is displayed in Fig.~\ref{SEDplot} as a vertical line.

\subsection{Incorporating these changes in introductory level classes}

We find that misunderstanding by the students often results when textbook authors and instructors who cover blackbody radiation introduce only $B_\lambda$ and the corresponding form of Wien's displacement law.  When students in these courses later encounter $B_\nu$ they may discover that they don't understand the Planck function as well as they thought they did.  And the reason for this, ultimately, is because the first class gave them an incomplete discussion.  

For science majors' classes, we think that, in the long run, the education of students will be enhanced by the introduction of both $B_\lambda$ and $B_\nu$ along with the Planckian FBSED.  The coverage of the material should include an explanation about why $B_\lambda$ and $B_\nu$ are different, that in order to spread the light into a spectrum one {\it must} choose a dispersion rule, such as spreading the light by frequency or by wavelength, and that this choice changes the overall shape.  The instructor can then introduce the Planckian FBSED as a type of plot designed for plotting the entire electromagnetic spectrum in a way that preserves the relative resolution in the entire plot, regardless of the method by which the spectrum is obtained (whether as a function of frequency or wavelength).

In non-science majors' classes, the Planck function is generally introduced in concept only, with a justified avoidance of presenting the analytical expression.  In the same vein, we recommend that instructors and textbook authors introduce the Planckian SED conceptually and to show {\it only} the Planckian FBSED.  Just as plots of 
$\log B_\lambda$ vs.\ $\log \lambda$ are currently shown with little explanation of the axes, plots of the Planckian FBSED can be given instead.  There is, really, no point in showing to this audience the other ways of displaying the Planck function.  The students need not be asked to comprehend the parameters on the axes any more than they are currently.

We suggest that Wien's displacement law be de-emphasized in general and not presented at all in introductory classes.  The equations for the average photon energy [Eq.~(\ref{enavg})], for the corresponding wavelength [Eq.~(\ref{lambdaavg})], for the total number of photons emitted per second per area [Eq.~(\ref{Nphotons})], and for the total energy flux [i.e.\ Stefan-Boltzmann law, Eq.~(\ref{SBphot})] can and should be presented.  Examples showing how the average photon energy depends on temperature would naturally replace examples done today with Wien's displacement law.  The students could also be taught that temperatures for thermal sources can be determined from fitting the Planck function to the measured intensity at a specific wavelength or to ratios of measured fluxes at different wavelengths; they should not be told that temperatures are inferred from fitting the location of the peak.  There should be no example that tries to explain the apparent color of the Sun.

In higher level classes, though, it might be suitable to introduce Wien's displacement law once the
students understand the significance of the dispersion rule used to create the spectrum under consideration.

\section{Conclusions}

Since a spectrum can be defined by spreading the light in various ways, such as into bins of either equal frequency or equal wavelength, and these will produce spectra with significantly different shapes and peak locations, simple and direct inferences from analysis of such spectra can be misleading.  Wien's displacement law, in particular, has been shown to be ambiguous, and not of practical use at the basic level.  Therefore, it should not be discussed in introductory classes.

We recommend that instructors and textbook authors who introduce blackbody radiation consider it standard practice to introduce the Planck function as a spectral energy per fractional bandwidth distribution 
($\nu B_\nu$ vs.\ $\ln\nu$ or $\lambda B_\lambda$ vs.\ $\ln\lambda$), that the average photon energy be presented in lieu of Wien's displacement law, and that the Stefan-Boltzmann law be presented in terms of both the total number of photons and total energy emitted per second per unit area.

\begin{acknowledgments}
The preparation of this paper was aided significantly by numerous very helpful discussions with Mark Heald, Sean Stewart, Leo Fleishman, and Michel Fioc.
\end{acknowledgments}

\newpage
\section*{Figures}

\begin{figure}[h]
%\scalebox{1.2}{\includegraphics{Marr_and_WilkinFig01.eps}}
\caption{\label{ThreeStarsB} The relative-magnitude spectra of $B_\lambda$ at temperatures of 5800~K (solid, black curve), 30,000~K (dashed, blue curve), and 4000~K (dotted, red curve) are shown.  The logarithmic brightness at different wavelengths is shown on a scale similar to that of stellar magnitudes.   The curves represent plots of 
$2.5 \log[{B_\lambda(\lambda) / B_\lambda(\lambda_{\textrm{ref}})}]$ vs.\ $\lambda / 1\,\textrm{nm}$, where $\lambda_{\textrm{ref}}$ is 400, 700, and 633 nm, respectively.  The $\lambda_{\textrm{ref}}$ for each curve corresponds to the location of the maximum of $B_\lambda$ in the visible band. 
For convenience, the vertical axis is plotted with the brighter values at the top.}
\end{figure}

\begin{figure}[h]
%\scalebox{1.2}{\includegraphics{Marr_and_WilkinFig02.eps}}
\caption{A plot of the spectral energy per fractional bandwidth distribution of blackbody radiation.  The parameter plotted on the abcissa is $\ln x$ where $x = h \nu / kT$ and the parameter on the ordinate is given by Eq.~(\ref{SEDva2}) divided by $T^4$, where $T$ is the temperature of the body.  To obtain $\nu B_\nu$, 
the values on the ordinate must be multiplied by $T^4$.  The vertical line represents the average photon energy, which is given by 
$\ln (x_{\langle E\rangle}) = 0.9937$ [see Eq.~(\ref{avgx})].  The peak of the curve, when displayed in this form, occurs at $\ln (x_{\textrm peak}) = \ln(3.9207) = 1.3663$.}
\label{SEDplot}
\end{figure}


\begin{thebibliography}{5}
\bibitem{Vega}Unlike standard astronomical practice, our relative magnitudes are not normalized to zero for Vega, which obscures the wavelength dependence of relative fluxes.
\end{thebibliography}
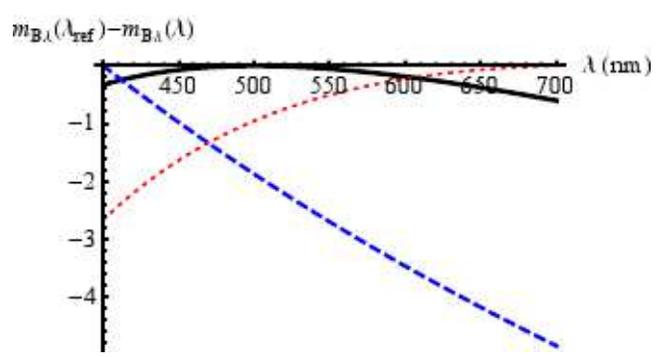
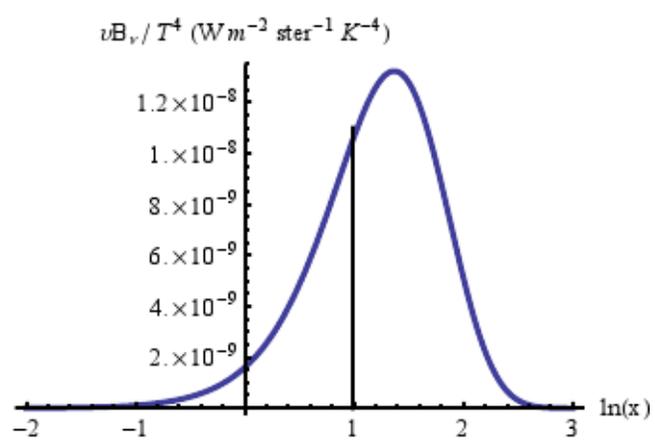
\end{document}